\documentclass[aps,prd,preprint,superscriptaddress,showpacs]{revtex4}

\usepackage{graphicx}
\usepackage{amsmath}
\usepackage{bm}
\usepackage{amssymb,enumerate}

\def\ee{\end{eqnarray}}

\def\=:{=\hspace{-.7em}\raisebox{1.1ex}{.}\hspace{.1em}\raisebox{-0.2ex}{.} }

\def\ee{\end{eqnarray}}

\def\=:{=\hspace{-.7em}\raisebox{1.1ex}{.}\hspace{.1em}\raisebox{-0.2ex}{.} }

\newcommand {\beq}{\begin{eqnarray}}
\newcommand {\eeq}{\end{eqnarray}}
\newcommand {\non}{\nonumber\\}

\newcommand {\1}[1]{\frac{1}{#1}}

\newcommand {\ph}{\varphi}

\newcommand {\del}{\partial}

\newcommand {\tr}{{\rm tr}\,}

\begin{document}


\title{Incarnations of Instantons 
}


\author{Muneto Nitta}

\affiliation{
Department of Physics, and Research and Education Center for Natural 
Sciences, Keio University, Hiyoshi 4-1-1, Yokohama, Kanagawa 223-8521, Japan\\
}


\date{\today}
\begin{abstract}
Yang-Mills instantons in a pure Yang-Mills theory  in four Euclidean space 
can be promoted to particle-like topological solitons 
in $d=4+1$ dimensional space-time. 
When coupled to Higgs fields, 
they transform themselves in the Higgs phase into Skyrmions, lumps and sine-Gordon kinks,  
with trapped inside a non-Abelian domain wall, 
non-Abelian vortex and monopole string, respectively. 
Here, we point out
that a closed monopole string, 
non-Abelian vortex sheet and 
non-Abelian domain wall 
in $S^1$, $S^2$ and $S^3$ shapes,  respectively, 
are all Yang-Mills instantons 
if their $S^1$, $S^2$ and $S^3$ moduli,  respectively,  
are twisted along their world-volumes.

\end{abstract}

\maketitle

\section{Introduction}

Recent discovery of non-Abelian vortices 
\cite{Auzzi:2003fs,Hanany:2003hp} 
and non-Abelian domain walls 
\cite{Shifman:2003uh,Eto:2005cc,Eto:2008dm} 
has been revealing relations among 
different topological solitons in diverse dimensions 
\cite{Eto:2005sw,Eto:2006pg,Shifman:2007ce}. 
When a 't Hooft-Polyakov monopole \cite{'tHooft:1974qc} 
is put into the Higgs phase, 
the magnetic fluxes from it are squeezed into the form of 
magnetic vortices,
becoming a confined monopole \cite{Tong:2003pz,Shifman:2004dr,Hanany:2004ea,Nitta:2010nd}. 
This configuration can be regarded as a kink inside 
a vortex. 
In other words, the monopole turns to the kink 
when it resides in the vortex. 
In the Higgs phase, Yang-Mills instantons are unstable to shrink 
in the bulk. However, they can stably exist as 
lumps (or sigma model instantons) 
\cite{Polyakov:1975yp} 
when trapped inside a non-Abelian vortex \cite{Hanany:2004ea,Eto:2004rz},
while they can stably exist as Skyrmions when 
trapped inside a non-Abelian domain wall 
\cite{Eto:2005cc}.
Recently, it has been also found that 
instantons transform themselves into sine-Gordon kinks when 
trapped inside a monopole string 
in a certain situation \cite{Nitta:2013cn}. 
Vortices or lumps become sine-Gordon kinks 
\cite{Nitta:2012xq,Kobayashi:2013ju,Jennings:2013aea} 
when trapped inside a ${\mathbb C}P^1$ domain wall 
\cite{Abraham:1992vb,Arai:2002xa}.
Skyrmions become lumps or 
baby Skyrmions \cite{Piette:1994ug,Weidig:1998ii}
inside a non-Abelian domain wall 
\cite{Nitta:2012wi}, and more generally 
$N$ dimensional Skyrmions become 
$N-1$ dimensional Skyrmions inside a non-Abelian domain wall 
\cite{Nitta:2012rq}.
Among those, one of the most successful applications 
to field theory may be made by  
confined monopoles \cite{Shifman:2004dr,Hanany:2004ea} 
which explain the coincidence of BPS spectra 
in 3+1 and 1+1 dimensions \cite{Dorey:1998yh}.

The ${\mathbb C}P^1$ model in $d=1+1$ dimensions 
offers a toy model of Yang-Mills theory in $d=3+1$ dimensions.
The presence of instantons \cite{Polyakov:1975yp} 
is one of such similarities between them.  
Sigma model instantons can be promoted to lumps in $d=2+1$ dimensions.
With a mass term admitting two vacua, 
the ${\mathbb C}P^1$ model allows 
a domain line with a $U(1)$ modulus 
\cite{Abraham:1992vb,Arai:2002xa}. 
If the $U(1)$ modulus winds around a straight domain line, 
a sine-Gordon kink is formed on it  corresponding 
to a lump in the bulk \cite{Nitta:2012xq},
which is a lower dimensional analogue of trapped instantons. 
The dynamics of such domain wall Skyrmions were 
studied \cite{Jennings:2013aea}.

\begin{table}[h]
\begin{tabular}{|cc|c|c|c|c|c|c|} \hline
& host solitons & bulk dim      &  codim. &  moduli &  w.v. shape &  w.v. soliton & homotopy\\ \hline\hline
(a) & ${\mathbb C}P^1$ domain wall& $2+1$ & $1$ & $S^1$ & ${\mathbb R}^1$ or $S^1$ & SG kink & $\pi_1(S^1)$\\ \hline
(b) & NA domain wall& & $1$ & $S^3$ & ${\mathbb R}^3$ or $S^3$ & Skyrmion & $\pi_3(S^3)$ \\
(c) &NA vortex sheet & $4+1$ & $2$  & $S^2$ &  ${\mathbb R}^2$ or $S^2$ & lump &$\pi_2(S^2)$\\
(d) & monopole string & & $3$  & $S^1$ &   ${\mathbb R}^1$ or $S^1$ & SG kink & $\pi_1(S^1)$\\ \hline
\end{tabular}
\caption{Host solitons of 
trapped instantons in $d=2+1$ (a) and in $d=4+1$ (b)--(d). 
The shape of the world-volume can be noncompact ${\mathbb R}^n$
or compact $S^n$, corresponding to 
trapped and untrapped instantons, respectively. 
NA denotes ``non-Abelian."
\label{table:instantons}}
\end{table}
\begin{figure}[h]
\begin{center}
\begin{tabular}{cc}
\includegraphics[width=0.48\linewidth,keepaspectratio]{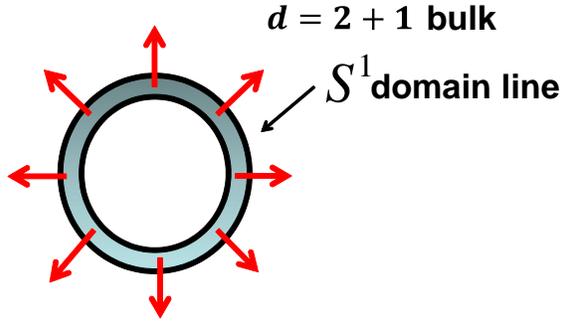} &
\includegraphics[width=0.50\linewidth,keepaspectratio]{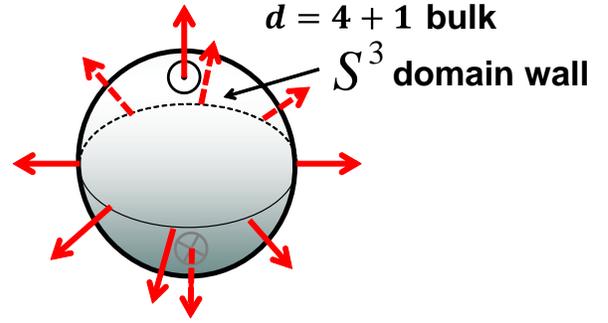} \\
(a) twisted closed domain line in $d=2+1$ & (b) twisted closed domain wall  in $d=4+1$ \\
\includegraphics[width=0.50\linewidth,keepaspectratio]{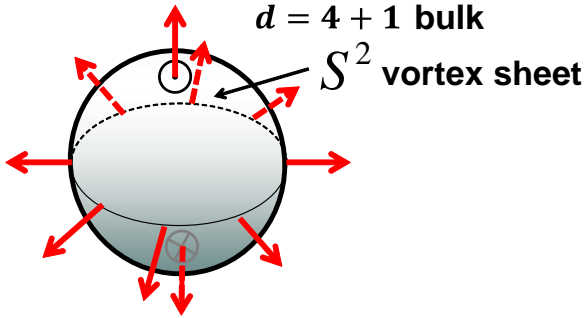} &
\includegraphics[width=0.50\linewidth,keepaspectratio]{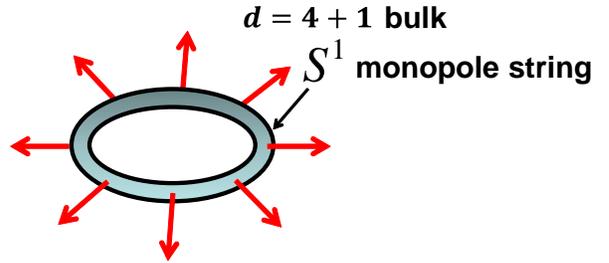}
\\
(c) twisted closed vortex sheet  in $d=4+1$& 
(d) twisted closed monopole string  in $d=4+1$
\end{tabular}
\end{center}
\caption{Incarnations of instantons in $d=2+1$ (a) and 
in $d=4+1$ (b)--(d). 
(a)--(d) correspond to those of Table \ref{table:instantons}.
The arrows schematically denote points in the $S^1$, $S^2$ and $S^3$ moduli.
\label{fig:instantons}
}
\end{figure}

Then, a question arises. 
Do all instantons have to be trapped into some host solitons 
to become composite states in theories in 
the Higgs phase?  
The answer is no. 
If one makes a closed domain line with the $U(1)$ 
modulus wound along it, 
such a twisted closed domain line
is nothing but an isolated lump \cite{Nitta:2012kj}.
This is stabilized against shrinkage and becomes 
a baby Skyrmion 
if one adds a four derivative Skyrme term 
in the original theory \cite{Kobayashi:2013ju}. 
An alternative way to stabilize a twisted closed domain line 
is to give a linear time dependence on the $U(1)$ modulus,
which results in a Q-lump \cite{Leese:1991hr}.
This situation is summarized in 
Table \ref{table:instantons} (a) and is illustrated in Fig.~\ref{fig:instantons} (a).
Thus, we have both trapped and untrapped instantons,
where the domain line world-volume is ${\mathbb R}^1$ and 
$S^1$, respectively. 
We may call the latter as incarnations of instantons.

Here, we propose higher dimensional analogues of this phenomenon 
for Yang-Mills instantons 
in Yang-Mills-Higgs theories  
in $d=4+1$ dimensions. 
In this dimensionality, instantons are particle-like solitons, 
while a vortex and monopole are a sheet (membrane) and string, 
respectively.  
In order to demonstrate our idea, 
we take the gauge group as 
$U(2) = {[SU(2) \times U(1)] / {\mathbb Z}_2}$ 
but generalizations to $U(N)= {[SU(N) \times U(1)] / {\mathbb Z}_N}$ 
or other groups are straightforward, 
since vortices in arbitrary gauge groups 
\cite{Eto:2008yi} 
such as $SO(N)$ and $USp(2N)$
 \cite{Eto:2009bg}
were already constructed.
We put the system into the Higgs phase 
where the $U(2)$ gauge group 
is spontaneously broken completely,  
by introducing some doublet Higgs fields with the common $U(1)$
charges. 
Unlike the case of $d=2+1$, there are three possibilities 
of incarnations of instantons. 
In Table \ref{table:instantons} (b)--(d), 
we summarize host solitons with world-volume ${\mathbb R}^{n,1}$ 
in which instantons can reside stably, {\it i.e.},  
a non-Abelian domain wall ($n=3$),  
non-Abelian vortex sheet  ($n=2$), 
and monopole string ($n=1$) 
of codimensions one, two and three, respectively. 
These solitons have internal 
moduli $S^3$, $S^2$ and $S^1$, respectively, 
as localized Nambu-Goldstone zero modes 
in addition to translational moduli.
When these moduli wind in the spatial world-volumes ${\mathbb R}^n$ 
of the host solitons according to 
the homotopy groups $\pi_n(S^n) \simeq {\mathbb Z}$, 
there appear Skyrmions, lumps and sine-Gordon kinks 
in the world-volumes of the domain wall, vortex sheet, and monopole string, 
respectively.  
They are all Yang-Mills instantons in the bulk point of view. 
Here, we have the following relation 
among the dimensionality and the number of moduli:
\beq
\# \mbox{(codim)} + \# \mbox{(moduli)} = \mbox{spatial dim (bulk)}, 
\eeq
for both $d=2+1$ and $4+1$ dimensions.

In this paper, we point out that when world-volumes of 
the host solitons are closed as $S^n$ instead to be flat and infinite,  
they can be regarded as incarnations of instantons,   that is,
untrapped instantons in the bulk, if the moduli $S^n$ wind along 
the world-volumes $S^n$,  
as in Fig.~\ref{fig:instantons}. 
Since the world-volumes are closed, 
the total topological charge of the host soliton 
is canceled out, and there remains only the instanton charge.
We need higher derivative terms for the stability of 
these solitons  in the same spirit with the Skyrme model,
which has been demonstrated explicitly 
for a twisted closed domain line as a baby Skyrmion 
in $d=2+1$ \cite{Kobayashi:2013ju}.

They are all higher dimensional generalizations of vortons. 
When a vortex string has a $U(1)$ modulus in $d=3+1$ dimensions, 
one can consider a vorton, 
which is a closed vortex string 
with the $U(1)$ modulus twisted along the string 
\cite{Davis:1988jq,Vilenkin:2000,Radu:2008pp}.
To enhance the stability, one usually considers  
a linear time dependence on  the $U(1)$ modulus.
The stability of the vorton in $d=3+1$ dimensions 
has been a longstanding problem  for decades after the proposal, 
and has been established recently \cite{Garaud:2013iba}.

This paper is organized as follows.
In Sect.~\ref{sec:models}, we introduce three models 
of $U(2)$ gauge theories coupled with some 
Higgs doublets. 
In Sect.~\ref{sec:trapped-instantons}, 
we present instantons trapped 
into host solitons, 
a domain wall, vortex-sheet 
and monopole-string. 
In Sect.~\ref{sec:untrapped-instantons}, 
we point out that 
closed host solitons 
are instantons when 
moduli are twisted along their world-volumes.
Sect.~\ref{sec:summary} 
is devoted to summary and discussion. 

\section{Yang-Mills-Higgs Theories}\label{sec:models}

We consider the following three models, 
$U(2)$ gauge theories 
coupled with  $N_{\rm F}$ doublet Higgs fields 
with the common $U(1)$ charge 
in $d=4+1$ dimensions. 
The field contents are 
the $U(2)$ gauge field $A_A$ $(A,B=0,1,2,3,4)$,  
a two by two real matrix of adjoint scalar fields $\Sigma$, 
and  a $2 \times N_{\rm F}$ 
matrix of complex scalar fields $H$.
The Lagrangians which we consider are of the form
\beq
&& {\cal L} = - \1{4 g^2}\tr F_{AB}F^{AB} 
 + \1{2g^2} \tr (D_{A} \Sigma)^2 
  + \tr D_{A}H^\dagger D^{A}H - V ,
\eeq
where $g$ is the gauge coupling, and the covariant derivatives are 
$D_{A}\Sigma = \del_A \Sigma -ig [A_A ,\Sigma]$ 
and 
$D_{A}H = \del_A H -ig A_A H$.
We consider the following three models 
in order to trap instantons into  
a domain wall, vortex, and monopole string.

\begin{enumerate}
\item
The theory 1 contains 
four Higgs doublets with the common $U(1)$ charge  
($N_{\rm F}=4$) summarized as 
a two by four matrix $H$. 
The potential term is (see,~e.g.,Ref.~\cite{Eto:2006pg})
\beq
&& V = g^2 \tr (H H^\dagger -v^2{\bf 1}_2)^2 
 + \tr \left[|\Sigma H - H M|^2 \right],
 \label{eq:Lagrangian1}
\eeq
with the mass matrix $M = {\rm diag.} (m,m,-m,-m)$.
The flavor symmetries are 
$SU(2)_{\rm L} \times SU(2)_{\rm R} \times U(1)$  
acting on the first and the last two flavors, respectively,  
for $m \neq0$, 
and $SU(4)$ for $m=0$.

\item
The theory 2 is consisting of 
two Higgs doublets with the common $U(1)$ charge 
($N_{\rm F}=2$) summarized as 
a two by two matrix $H$.
The potential term is (see,~e.g.,Ref.~\cite{Eto:2006pg})
\beq
&& V = g^2 \tr (H H^\dagger -v^2{\bf 1}_2)^2 .
\label{eq:Lagrangian2}
\eeq
We do not introduce mass parameters.
The flavor symmetry is $SU(2)$.

\item
The theory 3 contains 
two Higgs doublets with the common $U(1)$ charge 
($N_{\rm F}=2$) in 
a two by two matrix $H$.
The potential term is \cite{Nitta:2013cn}
\beq
&& V = g^2 \tr (H H^\dagger -v^2{\bf 1}_2)^2 
 + \tr \left[H(\Sigma - M)^2 H^\dagger\right] 
 + {\beta^2 \over v^2}\tr (H\sigma_x H^\dagger),
 \label{eq:Lagrangian3}
\eeq
with the mass matrix $M = {\rm diag.} (m,-m)$.
The case $m=\beta=0$ becomes the theory 2.
The flavor symmetry is $U(1)$ for $\beta=0,m\neq 0$ 
and no flavor symmetry for  $\beta\neq 0,m\neq 0$. 
\end{enumerate}

These Lagrangians 
can be made ${\cal N}=2$ supersymmetric 
({\it i.e.}, with eight supercharges)
by suitably adding fermions 
except for the case of $\beta \neq 0$ in the theory 3. 
The constant $v^2$ giving a VEV to $H$ is called 
the Fayet-Iliopoulos parameter 
in the context of supersymmetry.
In the limit of vanishing $v^2$, 
the systems go to the unbroken phase of the gauge group
where $H$'s decouple in the vacuum.

\section{Instantons inside host solitons with flat world-volume}\label{sec:trapped-instantons}
\subsection{Theory 1: Instantons inside a flat domain wall}
In the massless case $m=0$, 
the vacuum can be taken to be
\beq
 H 
= 
\left(
 v {\bf 1}_2 ,{\bf 0}_2 
\right) ,
\quad
\quad \Sigma ={\bf 0}_2
\eeq 
by using the flavor symmetry.
This is the so-called color-flavor locked vacuum.
The moduli space of vacua is 
the Grassmannian manifold $Gr_{4,2}\simeq SU(4)/[SU(2)\times SU(2)\times U(1)]$, 
see, e.g., Ref.~\cite{Higashijima:1999ki}.
In the massive case, $m \neq 0$, 
the vacuum is split into three disjoint vacua 
\beq
&& 
H = 
\left(
 v {\bf 1}_2 ,{\bf 0}_2 
\right) ,
\quad
\left(
\begin{array}{cccc}
 v &0 &0 &0 \\
 0 &0 &0 &v
\end{array}\right) ,\quad
{\rm or}
\quad 
\left(
  {\bf 0}_2 , v{\bf 1}_2 
\right) , 
\quad {\rm and} \quad \Sigma =v{\bf 1}_2,
\eeq 
with the following unbroken symmetries, respectively:
\beq
SU(2)_{\rm C+L} ,
\quad U(1) \times U(1),
\quad 
 {\rm or} \quad 
 SU(2)_{\rm C+R}.
\eeq

A non-Abelian domain wall solution 
interpolating between the first and third vacua, 
which is perpendicular to the $x^4$ coordinate, 
is given by \cite{Isozumi:2004jc,
Shifman:2003uh,Eto:2005cc,Eto:2008dm}
\beq
&& H_{\rm wall,0} = \1{\sqrt {1+|u_{\rm wall}|^2}}
      \left({\bf 1}_2, u_{\rm wall} {\bf 1}_2\right),
\quad
u_{\rm wall}(x^4) = e^{\mp m (x^4-X) + i \ph} , 
  \label{eq:wall-sol0}\\
&& \Sigma_{\rm wall,0} = v^{-2}H M H^\dagger ,
\quad 
A_{4,{\rm wall},0} = i v^{-2} (H\del_4 H^\dagger - \del_4H \cdot H^\dagger),
\nonumber
\eeq
in the strong gauge coupling limit. 
The general solution is 
\beq 
&& H_{\rm wall} = V H_{\rm wall,0} 
\left(\begin{array}{cc}
 V^\dagger & 0 \\ 0 & V
\end{array} \right) 
=  \1{\sqrt {1+|e^{\mp m (x^4-X) }|^2}}
      \left({\bf 1}_2, e^{\mp m (x^4-X) }U \right), 
\non
&&
 \Sigma_{\rm wall} = V \Sigma_{\rm wall,0} V^\dagger ,
\quad
A_{4,{\rm wall}} = V A_{4,{\rm wall},0} V^\dagger ,
\eeq 
with $V \in SU(2)$ and 
$U \equiv V^2 e^{i\ph} \in U(2)$.
Therefore the domain wall has 
${\mathbb R} \times U(2)$ moduli.
By using the moduli approximation \cite{Manton:1981mp,Eto:2006uw},
let us construct the effective theory of the domain wall 
with promoting the moduli $X$ and $U$ to fields 
 $X(x^i)$ and $U(x^i)$, respectively ($i=0,1,2,3$) 
on the world volume of the domain wall \cite{Shifman:2003uh,Eto:2005cc,Eto:2008dm}:
\beq
 {\cal L}_{\rm wall} = - {v^2 \over 4m} 
\tr \left(U^\dagger \del_{i} U
            U^\dagger \del^{i} U \right) 
+ {v^2 \over 2m} \del_i X \del^i X .
\eeq
By substituting the solution to gauge kinetic term, 
we have the Skyrme term \cite{Eto:2005cc}
\beq
 {\cal L}_{\rm wall}^{(4)} =  c \tr \left[ U^\dagger \del_i U,  U^\dagger \del_j U\right]^2 
\eeq
with a numerical constant $c$.

Since $\pi_3 [U(3)] \simeq {\mathbb Z}$, one can 
construct Skyrmions on the domain wall.  
One can confirm that 
Skyrmion solutions in the domain wall effective theory 
are instantons in the bulk \cite{Eto:2005cc}.

\subsection{Theory 2: Instantons inside a flat vortex sheet}
The system is in the unique color-flavor locked vacuum 
\beq
H = v {\bf 1}_2, \quad \Sigma={\bf 0}_2 ,
\eeq
where the unbroken symmetry is 
the color-flavor locked symmetry 
$SU(2)_{\rm C+F}$.
This model admits a non-Abelian $U(2)$ vortex solution \cite{Hanany:2003hp}, 
$H = {\rm diag.}\, (f(r) e^{i\theta}, v)$, 
where $(r,\theta)$ are polar coordinates in the $x^3$-$x^4$ plane, where the vortex world-volume has the 
coordinates $(x^0,x^1,x^2)$.
The transverse width of the vortex is $1/gv$. 
The vortex solution breaks the vacuum symmetry 
$SU(2)_{\rm C+F}$ into $U(1)$ in the vicinity of the vortex, 
and consequently 
there appear ${\mathbb C}P^1 \simeq U(2)_{\rm C+F}/U(1)$ 
Nambu-Goldstone modes localized around the vortex. 
The vortex solutions have the orientational moduli 
 ${\mathbb C}P^1$ in addition to the translational (position) moduli $Z$.  
By promoting the moduli to the fields depending on 
the world-volume coordinates $(x^0,x^1,x^2)$, 
the low-energy effective theory of these modes 
can be constructed to yield 
the ${\mathbb C}P^1$ model in  $d=2+1$ dimensional 
vortex effective theory \cite{Hanany:2003hp,Auzzi:2003fs,Tong:2003pz,Shifman:2004dr,
Eto:2004rz,Eto:2006uw}  
($\mu=0,1,2$)
\beq
&& {\cal L}_{\rm vort.eff.}
=  2 \pi v^2 |\del_{\mu} Z|^2 + {4\pi \over g^2} \left[ 
 {\partial_{\mu} u^* \partial^{\mu} u 
  \over (1 + |u|^2)^2} \right] .\label{eq:vortex-th1}
\eeq
Here $Z(x^{\mu}), u(x^{\mu}) \in {\mathbb C}$ represent 
the position and orientational moduli 
(the projective coordinate of ${\mathbb C}P^1$), respectively.
The points $u=0$ and $u=\infty$ corresponding to 
the north and south poles of the target space 
${\mathbb C}P^1$. 

Since $\pi_2({\mathbb C}P^1) \simeq 
\pi_2(S^2) \simeq {\mathbb Z}$, lumps can exist 
on the vortex world-volume. 
The BPS equation for lumps in the ${\mathbb C}P^1$ model 
(with $z \equiv x^1+ i x^2$) and the solutions are
\beq
&& \del_{\bar z} u = 0 ,\\
&& u_{\rm lump} = \sum_{a=1}^k {\lambda_a \over z-z_a}
\eeq
respectively. 
The lump charge and energy are 
\beq
&& T_{\rm lump} 
\equiv \int d^2x {i (\partial_i u^* \partial_j u - \partial_j u^* \partial_i u )
\over (1+|u|^2)^2} = 2 \pi k ,\\
&& E_{\rm lump} = {4\pi \over g^2} T_{\rm lump} 
= {4\pi \over g^2} \times 2\pi k 
= {8\pi^2 \over g^2} k = E_{\rm inst}, \label{eq:lump-inst}
\eeq
showing that the lumps inside the vortex are instantons in the bulk.
 
\begin{figure}[ht]
\begin{center}
\includegraphics[width=0.50\linewidth,keepaspectratio]{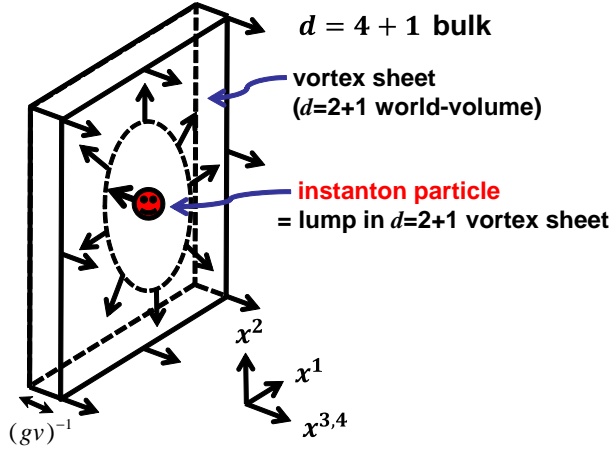}
\end{center}
\caption{An instanton trapped inside a non-Abelian vortex sheet 
in $d=4+1$. In the $2+1$ dimensional world-volume of the vortex sheet, 
it is realized as a ${\mathbb C}P^1$ lump. 
The arrows denote points in the $S^2$ moduli. 
\label{fig:trapped-instantons}
}
\end{figure}
\subsection{Theory 3: Instantons inside a straight monopole string}

In this case, we need two different host solitons, 
a vortex sheet and monopole string. 
We embed a configuration of a kink inside a monopole string 
into 
a non-Abelian vortex sheet. 
We start to deform the vortex theory in the previous subsection 
by mass $m$.
To this end, we consider the limit $\beta=0$ for a while. 
In the presence of mass,  $m\neq 0$, 
the $SU(2)_{\rm C+F}$ symmetry is explicitly broken 
and 
the system is in the color-flavor locked vacuum 
\beq
H = v {\bf 1}_2, \quad \Sigma= {\rm diag.}(m,-m). 
\eeq

Let us consider the non-Abelian vortex as in the theory 2.
Considering a regime $m \ll gv$ of small mass, 
it induces the mass in the $d=2+1$ dimensional 
vortex effective theory \cite{Hanany:2003hp,Tong:2003pz,Eto:2004rz,Eto:2006uw}  
($\mu=0,1,2$)
\beq
&& {\cal L}_{\rm vort.eff.}
=  2 \pi v^2 |\del_{\mu} Z|^2 + {4\pi \over g^2} \left[ 
 {\partial_{\mu} u^* \partial^{\mu} u - m^2 |u|^2 
  \over (1 + |u|^2)^2} \right] .\label{eq:vortex-th2}
\eeq
This is the massive ${\mathbb C}P^1$ model, 
which can be made supersymmetric with fermions 
\cite{Abraham:1992vb,Arai:2002xa}.

A monopole solution can be constructed as a domain wall interpolating the two vacua 
$u=0$ and $u=\infty$ \cite{Abraham:1992vb,Arai:2002xa} in the vortex effective theory (\ref{eq:vortex-th2}).
The solution 
\beq
 u_{\rm mono.}(x^2) = e^{\mp m (x^2-Y) + i \ph} , \label{eq:wall-sol}
\eeq
is placed perpendicular to the $x^2$-coordinate, 
where $\mp$ represents a monopole and an anti-monopole 
with the width $1/m$.
Here, $Y$ and $\ph$ are moduli parameters 
representing the position 
and $U(1)$ phase 
of the (anti-)monopole. 
The domain wall tension 
$E_{\rm wall}= {4\pi \over g^2} \times m = E_{\rm mono.}$
coincides with the monopole mass $E_{\rm mono.}$ 
and the monopole charge in the bulk theory,
showing that the wall in the vortex theory 
is a monopole string in  
the bulk \cite{Tong:2003pz}. 
The effective theory of the monopole string 
is obtained 
by promoting the moduli $Y$ and $\ph$ to fields 
 $Y(x^i)$ and $\ph(x^i)$ ($i=1,2$) \cite{Manton:1981mp,Eto:2006uw} 
on the string as \cite{Arai:2002xa}
\beq
 {\cal L}_{\rm mono.eff.} 
&=& {4\pi \over g^2} \1{2m} [(\del_i Y)^2 + (\del_i\ph)^2],
\eeq
which is a free theory, a sigma model with the target space 
${\mathbb R}\times U(1)$. 

Since $\pi_1[U(1)]\simeq {\mathbb Z}$ there can exist 
a phase kink on the monopole string.
However, a phase kink is unstable against the expansion and is diluted along the string.
It can be stabilized by turning on $\beta$ perturbatively 
($\beta \ll m v$),
which deforms the vortex effective Lagrangian 
by the potential term \cite{Eto:2009tr,Nitta:2013cn} 
\beq
\Delta {\cal L}_{\rm vort.eff.} =
 - c \beta^2  {u + u^*\over 1+|u|^2} ,  
\quad 
 c = \sqrt{2} \pi \int_{0}^{\infty} dr\, r (v^2 - f^2 ) 
\label{eq:Josephson}
\eeq
with the vortex profile function $f$, 
and 
the monopole effective Lagrangian by 
\cite{Nitta:2013cn}
\beq
 \Delta {\cal L}_{\rm mono.eff.} 
 = c \beta^2 \int_{-\infty}^{+\infty} dy 
 {e^{my+ i\ph} + e^{my- i\ph} \over 1+e^{2my}} 
 = {\pi c \beta^2 \over m} \cos \ph.
\eeq
We thus obtain the sine-Gordon model 
${\cal L}_{\rm mono.eff.} +  \Delta {\cal L}_{\rm mono.eff.}$ 
with the additional field $Y$.

\begin{figure}[ht]
\begin{center}
\begin{tabular}{cc}
\includegraphics[width=0.46\linewidth,keepaspectratio]{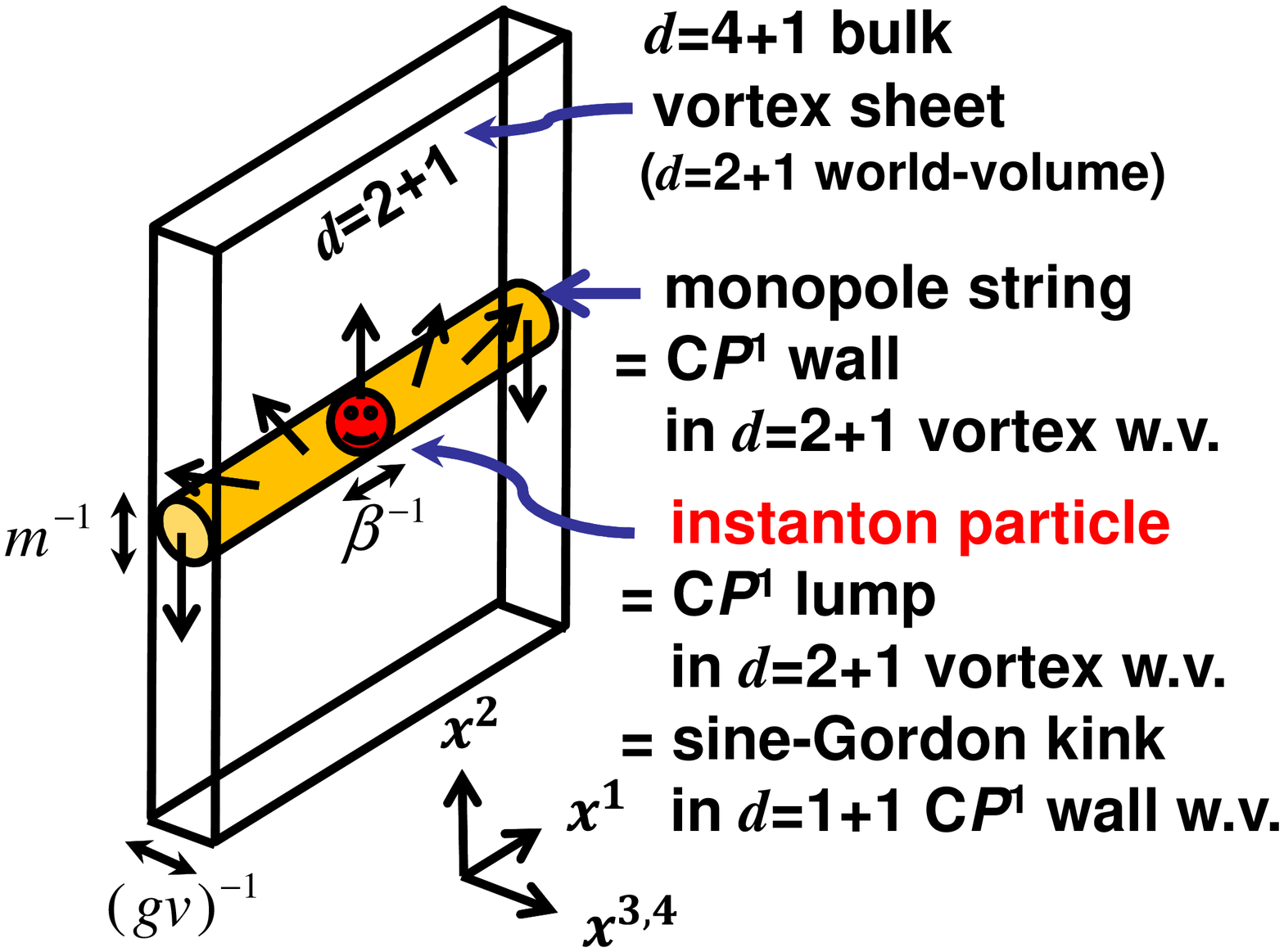}
&
\includegraphics[width=0.54\linewidth,keepaspectratio]{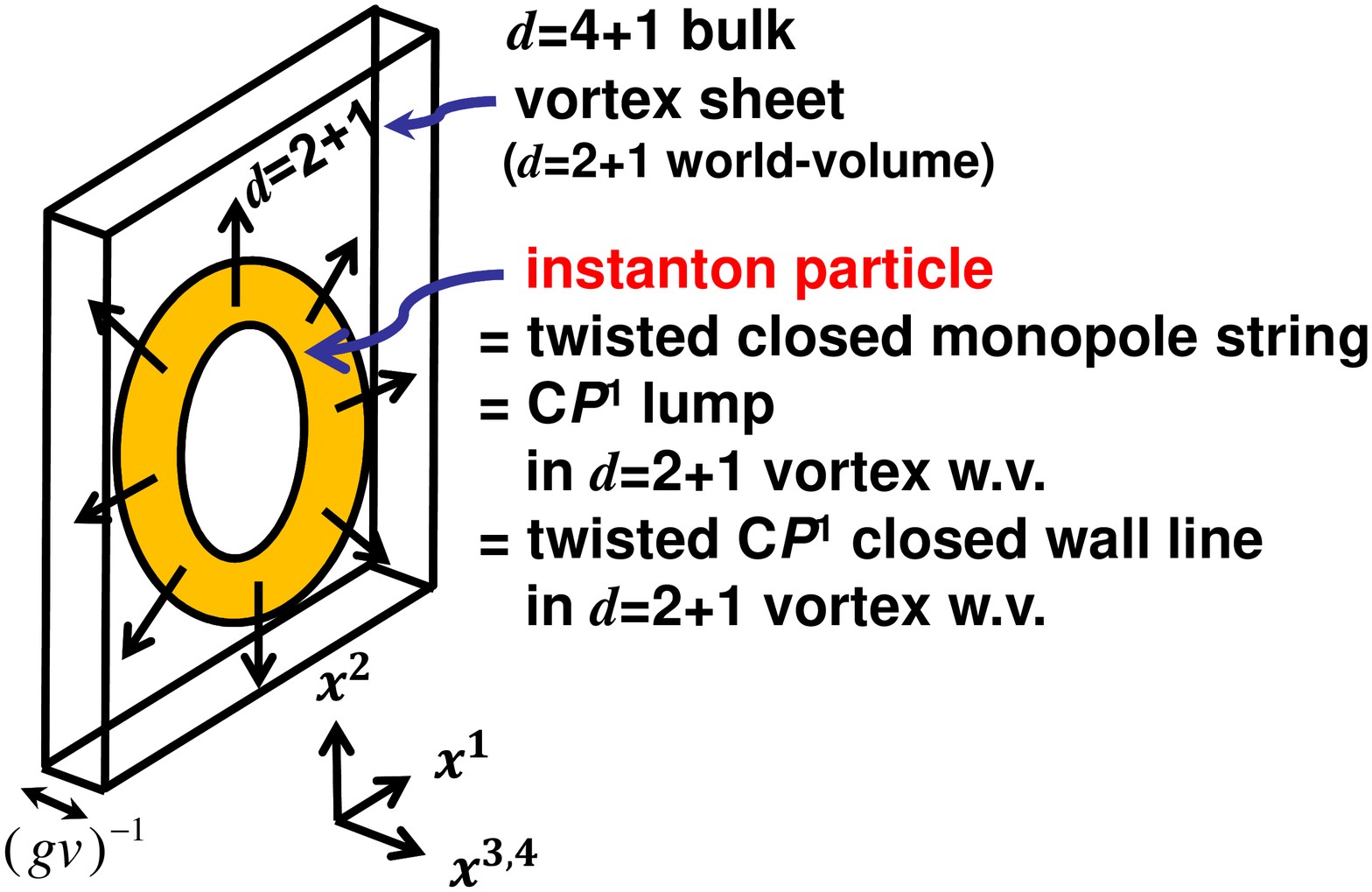} 
\\
(a) & (b)
\end{tabular}
\end{center}
\caption{(a) An instanton trapped inside a monopole string 
in $d=4+1$. In the $1+1$ dimensional world-volume of the monopole string, it is realized as a sine-Gordon kink. 
The arrows denote points in the $S^1$ modulus.  
(b) An instanton as twisted closed monopole string.
\label{fig:instanton-on-monopole}
}
\end{figure}

We construct an instanton as a sine-Gordon kink. 
The BPS equation and its a one-kink solution  are
\beq
&& \del_i\ph \pm \tilde \beta \sin {\ph\over 2} = 0, 
\quad \tilde \beta^2 \equiv \1{2} {\tilde c }\beta^2
\\
&& \ph = 4 \arctan \exp{{\tilde \beta \over 4} (x- X)} 
 + {\pi \over 2},\ \quad
\eeq
respectively, with the width $\Delta x \sim 1/\tilde \beta$.
The topological charge and 
energy are 
\beq
 T_{\rm SG} = 
{4\tilde \beta \over m}, \quad
 E_{\rm SG} =   {2 \pi \over g^2 m} T_{\rm SG} 
= {8 \pi \tilde \beta \over g^2 m^2} ,
  \label{eq:SG-energy}
\eeq
respectively. 
One can confirm that 
 $k$ sine-Gordon kinks can be identified with $k$ ${\mathbb C}P^1$ lumps with  
the topological charge $k \in \pi_2 ({\mathbb C}P^1)$ \cite{Polyakov:1975yp} 
in $d=2+1$ dimensional vortex world-volume,  
by explicitly calculating a lump charge
$T_{\rm lump} = 2 \pi k$ \cite{Nitta:2013cn}, 
and they can be further 
 identified with Yang-Mills instantons in the bulk \cite{Eto:2004rz} as in Eq.~(\ref{eq:lump-inst}). 
We conclude that the sine-Gordon kink on the monopole string 
corresponds to a Yang-Mills instanton in the bulk point of view, 
as schematically illustrated in Fig.~\ref{fig:instanton-on-monopole} (a). 
This has been checked by taking various limits \cite{Nitta:2013cn}.
For instance, in the limit $m,\beta \to 0$ with keeping 
$\beta/m^2$ fixed, the configuration becomes 
the instanton trapped inside a non-Abelian vortex sheet 
in Fig.~\ref{fig:trapped-instantons} 
in the theory 2.

\section{Instantons as twisted closed solitons}
\label{sec:untrapped-instantons}

\subsection{Theory 3: Instantons as twisted closed monopole strings}

First, we assume a closed monopole string 
as a background. 
By making a closed loop, the total monopole charge 
is canceled out. 
Since the closed monopole string has a finite world-volume,
we do not need to localize the phase kink 
to the form of a sine-Gordon kink.
Therefore, 
we set $\beta=0$ and the phase gradient becomes 
uniform along the closed monopole string 
as in Fig.~\ref{fig:instantons} (d). 
In order to stabilize the closed monopole string, 
there are two possible ways. 
One is giving a time dependence and the other 
is to add higher derivative terms in the Lagrangian 
in the same spirit with the Skyrme model. 

Let us discuss the first possibility. 
In this case, in fact, we do not need the help of 
the vortex, so we take the limit $v=0$ 
in which the vortex is diluted and eventually disappears.
In taking this limit, we keep $m \neq 0$ 
with non-vanishing VEVs for $\Sigma ={\rm diag}(m,-m)$. 
The closed monopole string 
with the $U(1)$ phase twisted along the string 
and the linear time dependence 
on the $U(1)$ phase is known as 
a dyonic instanton \cite{Lambert:1999ua}.
This is a BPS state and is stable 
in supersymmetric gauge theory.

This is a higher dimensional analogue of a vorton,   
which is a closed vortex string in $d=3+1$ dimensions 
with the $U(1)$ phase twisted along the string 
and the linear time dependence on the $U(1)$ phase.
The stability of such a soliton has been established recently 
\cite{Garaud:2013iba}. 

We can also consider the configuration inside a non-Abelian vortex
sheet for $v \neq 0$ and $m\neq 0$, 
see Fig.~\ref{fig:instanton-on-monopole}(b). 
The dyonic instanton inside a non-Abelian vortex 
is a twisted closed domain line with 
a linear time dependence on the $U(1)$ phase. 
This is nothing but a Q-lump \cite{Leese:1991hr}.
The Q-lump is a 1/2 BPS state in the massive ${\mathbb C}P^1$ model 
and the total configuration of the Q-lump inside the vortex 
is a 1/4 BPS state in supersymmetric gauge theories 
\cite{Eto:2005sw}.

Higher derivative corrections to the vortex effective 
theory was studied in Refs.~\cite{Eto:2012qda,Liu:2009rz}.
If we calculate the next-to-leading order 
to the moduli approximation, we obtain four derivative terms 
in the ${\mathbb C}P^{N-1}$ Lagrangian \cite{Eto:2012qda}. 
However, this form for the BPS vortices 
is not suitable for stabilizing the lumps 
in the vortex.  
Then, let us discuss the possibility 
of adding higher derivative terms 
in the original Lagrangian. 
The Lagrangian containing possible four derivative terms
can be written as 
\beq
&& {\cal L}^{(4)} 
= - {1 \over g'^2} \tr F_{AB}^3 
+ c_1 \tr (D_A \Sigma D^A \Sigma  D_B \Sigma D^B \Sigma)
+ c_2 [\tr (D_A \Sigma D^A \Sigma)]^2 \non
&& \qquad + d_1  \tr (D_A H^\dagger D^A H D_B H^\dagger D^B H)
+ d_2  [\tr (D_A H^\dagger D^A H)]^2 \label{eq:4-deriv}
+ O(\del^6).
\eeq
In order to stabilize the configuration, 
we need at least 
four derivative terms in Eq.~(\ref{eq:4-deriv}) 
in the original Lagrangian from the beginning.
These terms would induce more general four derivative terms in 
the vortex effective theory \cite{Liu:2009rz}.
Then, lumps become baby Skyrmions in ${\mathbb C}P^1$ model 
with four derivative term \cite{Piette:1994ug,Weidig:1998ii}.
With the mass term which we are considering, 
these lumps are in the form of twisted domain wall rings  \cite{Weidig:1998ii,Kobayashi:2013ju}. 
Only the first term contain the time derivatives in the second order;
the rests contain the fourth order time derivatives 
and may be unsuitable for stabilization of solitons.

In the end, let us consider 
the limit  $v \to 0$ in which the Higgs fields $H$  
decouple from other fields and the vortex 
disappears.  
We expect that 
the twisted closed monopole string remains stable 
without the help of the vortex 
in a certain parameter region 
because of the presence of the four derivative terms 
for the adjoint scalar fields $\Sigma$ in the first line in 
Eq.~(\ref{eq:4-deriv}).
This remains as a challenging future problem, 
if one notes that 
the stability of the vorton in $d=3+1$ dimensions 
has been a longstanding problem until recently 
\cite{Garaud:2013iba} for decades after the proposal \cite{Davis:1988jq}. 

\bigskip
In $d=3+1$ dimensions, one can interpret 
it as an instanton in the following way 
by identifying one axis in Fig.~\ref{fig:instantons}(d)
as the time direction. 
First, 
a pair of a monopole and an anti-monopole is created 
with the same $S^1$ moduli, say the down arrow.
Subsequently, 
the $S^1$ moduli of the monopole and anti-monopole gradually change in time 
clockwise and counterclockwise, respectively  
 toward the opposite point of $S^1$ the up arrow.
Finally, they annihilate each other with 
the same $S^1$ moduli.

\subsection{Theory 2: Instantons as twisted closed vortex sheets}

An instanton trapped in a non-Abelian vortex sheet 
with a flat and infinite world-volume ${\mathbb R}^{2,1}$ 
is schematically drawn in Fig.~\ref{fig:trapped-instantons}.
Since the infinities of the spatial world-volume ${\mathbb R}^2$ 
are identified, the configuration is topologically $S^2$ 
by one point compactification. 
We now physically compactify the world-volume ${\mathbb R}^2$ to 
$S^2$ by a stereographic map  
to obtain a twisted closed vortex sheet 
drawn in Fig.~\ref{fig:instantons} (c).
In this case, the host soliton, the non-Abelian vortex sheet, 
has $S^2$ moduli, for which a dyonic extension is impossible. 
Only possibility to stabilize the closed vortex sheet is 
to consider higher derivative terms in Eq.~(\ref{eq:4-deriv}). 

In $d=3+1$ dimensions, one can interpret 
this configuration as an instanton in the following way
by identifying the  vertical axis in Fig.~\ref{fig:instantons}(c)
as the time direction. 
First, a closed vortex string with 
the same $U(1)$ phase, say the south pole,  
is created at one point. 
Subsequently, the $S^2$ moduli gradually change 
directions along the closed string 
toward the opposite point of the $S^2$, the north pole.
Finally, the closed vortex string shrinks with 
the same $S^2$ moduli, the north pole, 
to be annihilated at one point.  

\subsection{Theory 1: Instantons as twisted closed 
domain walls}

Finally, we consider 
a non-Abelian domain wall having $S^3$ moduli.
When the domain wall has a flat world-volume ${\mathbb R}^{3,1}$,  
the infinities of the spatial world-volume ${\mathbb R}^3$ 
are identified, and the configuration is topologically $S^3$ 
by one point compactification. 
We physically compactify  ${\mathbb R}^3$ to $S^3$ 
and consider a Skyrmion on it 
 as in Fig.~\ref{fig:instantons} (b). 
Since the non-Abelian domain wall has the $S^3$ moduli,
a dyonic extension is again impossible. 
Only possibility to stabilize these solitons is 
to consider higher derivative terms in Eq.~(\ref{eq:4-deriv}).

We can interpret 
this configuration as an instanton
in $d=3+1$ dimensions 
by identifying the  vertical axis in Fig.~\ref{fig:instantons}(b)
as the time direction 
as before. 
After a closed domain wall of an $S^2$ shape 
with the same $S^3$ moduli 
is created at one point, 
it grows and shrinks at the other point, 
where the $S^3$ moduli wind as before. 

\section{Summary and Discussion}\label{sec:summary}
Although Yang-Mills instantons are usually unstable to shrink 
in the Higgs phase, 
they can stably live inside some host solitons, 
such as a domain wall, vortex sheet, or monopole string, 
as Skyrmions, lumps, or sine-Gordon kinks, 
respectively. 
We have pointed out 
instantons can exist as 
twisted closed domain walls, vortex sheet 
or monopole string, 
when world-volumes of these host solitons 
are compact, $S^3$, $S^2$, or $S^1$, respectively, 
and the moduli $S^3$, $S^2$, or $S^1$ are 
wound around the world-volume,   
as summarized in Table \ref{table:instantons} and 
Fig~\ref{fig:instantons}. 
Maps from world-volume of the host solitons 
to the moduli space of the host solitons are 
nontrivial; 
$\pi_3(S^3)$, $\pi_2(S^2)$ and $\pi_1(S^1)$ 
for a closed domain wall, vortex sheet 
and monopole string, respectively. 
We have called these solitons 
as incarnations of instantons. 
They are all higher dimensional generalizations 
of a vorton, vortex ring with the $U(1)$ modulus twisted. 
The stability of these solitons will need 
higher derivative terms as in Skyrmions 
in the Skyrme model. 
While the lower dimensional version 
of incarnation of instantons 
(twisted closed domain walls)
was constructed numerically
\cite{Kobayashi:2013ju}, 
construction of numerical 
solutions for twisted solitons 
in 4+1 dimensions   
remains as an important future problem.

A spherical domain wall with the $S^2$ moduli twisted 
as a Skyrmion 
has been numerically constructed 
in a modified Skyrme model 
\cite{Gudnason:2013qba}.
There, the conventional Skyrme term has been
taken to be negative and the sixth order term 
has been added, 
in order to have a stable spherical world-volume of 
the domain wall.
The same may be needed for spherical solutions 
in our case;  a gauge kinetic term 
$\tr F_{AB}^2$ 
with the negative sign
and a higher order term $\tr F_{AB}^3$.

Without higher derivative terms, 
all models which we considered 
can be made supersymmetric 
with eight supercharges 
(the term $\beta (\neq 0)$ in the theory 3 
breaks supersymmetry, which 
is needed  for the stability of sine-Gordon kink 
on a straight monopole-string, 
but that term is not needed for 
a closed monopole string, 
where supersymmetry is preserved).
Higher derivative terms may be added 
preserving supersymmetry  
which is a future problem.

In this paper, we have considered 
shapes of world-volumes are $S^n$.
One may also consider toroidal shape $T^n$ 
or other geometries. 
For instance, 
toroidal domain walls in $d=3+1$ admit two cycles 
along which the $U(1)$ modulus of the domain wall 
can be twisted 
with two winding numbers.
In this case, they carry a Hopf charge 
instead of the instanton charge, 
see, e.g., Refs.~\cite{Kobayashi:2013bqa}.
Relations between world-volume geometry 
and topological charge should be clarified.

There is a $d=2+1$ dimensional version, 
where a lump is a twisted domain wall 
as in Fig~\ref{fig:instantons} (a). 
There may be higher dimensional version 
of incarnation of instantons. 
For instance, 
six dimensional instantons 
\cite{Tchrakian:1978sf,Kihara:2007di} 
may be realized as 
an $S^5$ domain wall, $S^4$ vortex, 
$S^3$ monopole, and so on.

\section*{Acknowledgements}

This work is supported in part by 
a Grant-in-Aid for Scientific Research (No. 25400268) 
and by the ``Topological Quantum Phenomena'' 
Grant-in-Aid for Scientific Research 
on Innovative Areas (No. 25103720)  
from the Ministry of Education, Culture, Sports, Science and Technology (MEXT) of Japan. 


\end{document}